\documentclass[10pt,conference]{IEEEtran}
\IEEEoverridecommandlockouts

\usepackage{cite}
\usepackage{graphicx}
\usepackage{amsmath,amssymb,mathtools}
\usepackage{bm}
\usepackage{booktabs}
\usepackage{url}

\usepackage[left=0.673in,right=0.673in,top=0.747in,bottom=1.1in]{geometry}

\usepackage[draft]{hyperref}

\usepackage{array}
\usepackage{tabularx}
\usepackage{ragged2e}
\newcolumntype{L}[1]{>{\RaggedRight\arraybackslash}p{#1}}
\newcolumntype{Y}{>{\RaggedRight\arraybackslash}X}
\usepackage{algorithm}
\usepackage{algpseudocode}

\algrenewcommand\algorithmiccomment[1]{\hfill$\triangleright$~#1}

\allowdisplaybreaks

\DeclareMathOperator*{\argmax}{arg\,max}
\DeclareMathOperator*{\clip}{clip}

\begin{document}

\title{ARC: Alignment-based RPM Estimation with Curvature-adaptive Tracking}

\author{
\IEEEauthorblockN{Weiheng Hua, Changyu Hao}
\IEEEauthorblockA{
\textit{Blue Aspirations}\\
Hangzhou, China\\
\{weiheng.hua, changyu.hao\}@blue-aspirations.com
}
}

\maketitle

\begin{abstract}
Tacho-less rotational speed estimation is critical for vibration-based prognostics and health management (PHM) of rotating machinery, yet traditional methods—such as time-domain periodicity, cepstrum, and harmonic comb matching—struggle under noise, non-stationarity, and inharmonic interference. Probabilistic tracking offers a principled way to fuse multiple estimators, but a major challenge is that heterogeneous estimators produce evidence on incompatible axes and scales. We address this with ARC (Alignment-based RPM Estimation with Curvature-adaptive Tracking) by unifying the observation representation. Each estimator outputs a one-dimensional evidence curve on its native axis, which is mapped onto a shared RPM grid and converted into a comparable grid-based log-likelihood via robust standardization and a Gibbs-form energy shaping. Standard recursive filtering with fixed-variance motion priors can fail under multi-modal or ambiguous evidence. To overcome this, ARC introduces a curvature-informed, state-dependent motion prior, where the transition variance is derived from the local discrete Hessian of the previous log-posterior. This design enforces smooth tracking around confident modes while preserving competing hypotheses, such as octave alternatives. Experiments on synthetic stress tests and real vibration-table data demonstrate stable, physically plausible trajectories with interpretable uncertainty, and ablations confirm that these gains arise from uncertainty-aware temporal propagation rather than per-frame peak selection or ad hoc rules.
\end{abstract}


\section{Introduction}
Rotational speed (RPM) is a fundamental prerequisite for vibration-based diagnostics and prognostics of rotating machinery. Techniques such as order analysis and angular resampling require an accurate speed reference to decouple the spectral content from speed fluctuations \cite{Lu2019TIM_TacholessReview}. However, in industrial deployments, installing tachometers or encoders is often impractical due to cost, space constraints, and reliability concerns \cite{Lu2019TIM_TacholessReview,Peeters2019MSSP_ReviewComparison}. This motivates tacho-less speed estimation: recovering the time-varying RPM trajectory directly from vibration signals to enable downstream Prognostics and Health Management (PHM).

Extracting a robust speed trajectory from industrial vibration data is challenging. Many existing approaches rely on framewise estimators that derive local evidence from short-time snapshots, e.g., lag-domain periodicity costs (YIN) \cite{deCheveigne2002YIN}, quefrency-domain peaks (cepstrum), harmonic comb matching scores, or time--frequency ridge evidence \cite{Peeters2019MSSP_ReviewComparison}. Although effective under favorable conditions, such snapshot-based evidence can be fragile in the presence of strong interharmonics, resonance amplification, and broadband disturbances. A typical failure mode is octave ambiguity—locking onto a sub- or super-harmonic—which invalidates downstream order analysis.

To improve robustness, a natural idea is to exploit multiple cues over time \cite{Peng2023MSSP_AssessImprove,Yoo2024MSSP_WMOVA}. However, most existing pipelines combine information under a largely homogeneous representation. A common practice is to produce one speed estimate per frame (for example by selecting a ridge or a dominant candidate in a time--frequency map) and then enforce temporal continuity using a tracking or smoothing stage \cite{Feng2023SHM_VKF}. Another line constructs a single cost surface in a chosen analysis domain and extracts a globally consistent trajectory by dynamic programming or Viterbi-style search \cite{Yoo2024MSSP_WMOVA,Zhang2025MST_ARE,Leclere2016MOPA}. While effective for cues sharing a common coordinate system, these strategies struggle to fuse fundamentally different estimators—a practical necessity since no single representation is universally reliable. 
Such estimators output evidence on incompatible axes and with incomparable scales: time-domain methods produce cost curves on a lag axis, cepstrum produces peak prominence on a quefrency axis, and spectral or time--frequency methods yield scores on frequency- or speed-related coordinates. Without a principled way to reconcile these heterogeneous evidence curves, ad hoc normalization or collapsing them into a single confidence value may distort the relative evidence structure, suppress multi-modality, and lead to over-confident yet incorrect decisions, especially under octave ambiguity~\cite{Peeters2019MSSP_ReviewComparison}.

To address the above gap, we propose ARC (Alignment-based RPM estimation with Curvature-adaptive tracking) and first introduce its alignment component, termed C2G (Curve-to-Grid). Rather than outputting a single point estimate, each estimator produces a dense 1-D evidence curve on its native axis. C2G deterministically maps these curves onto a shared RPM grid, applies robust standardization to remove estimator-specific offsets and scales, and converts the aligned evidence into grid-based log-likelihoods via a Gibbs-form energy shaping. This unified RPM-grid representation enables principled multi-estimator fusion while preserving multi-modal hypotheses (e.g., competing octave candidates) for downstream decision-making. In this work, we instantiate the framework with three representative and complementary estimators—YIN, cepstrum, and comb matching—which provide evidence on distinct axes (lag/quefrency/frequency) and thus expose the core difficulty of heterogeneous fusion. Importantly, the C2G component is estimator-agnostic and can accommodate additional evidence sources.

Furthermore, to balance trajectory smoothness and responsiveness, ARC performs curvature-adaptive recursive tracking with a state-dependent motion prior. Instead of relying on a fixed-variance motion prior—which is often either too stiff (missing genuine speed changes) or too loose (inviting jitter and mode hopping)—we adapt the transition variance using the local curvature of the previous log-posterior, estimated via a discrete Hessian and motivated by a Laplace-approximation intuition. Consequently, the tracker automatically tightens around confident modes during steady operation and widens its search under ambiguous multi-modal evidence. This mechanism supports rapid recovery from octave traps and enables accurate tracking of abrupt speed changes without requiring ad hoc rules.

The main contributions of this article are summarized as follows:
\begin{itemize}
    \item \textbf{C2G Observation Alignment:} A systematic Curve-to-Grid (C2G) protocol that aligns heterogeneous 1-D evidence curves---originating from distinct domains such as lag, quefrency, and frequency---onto a shared RPM grid. This establishes a unified hypothesis space for fusing incompatible observations without collapsing multi-modality.
    \item \textbf{RPM-Grid Likelihood Construction:} A robust standardization and Gibbs-form energy shaping that converts aligned RPM-grid evidence into comparable RPM-grid log-likelihoods by removing estimator-specific offsets and scales, thereby avoiding arbitrary range normalization.
    \item \textbf{ARC Curvature-adaptive Recursive Tracking:} An online grid-based recursive tracker with a state-dependent motion prior whose transition variance is modulated by the local curvature of the previous log-posterior (estimated via a discrete Hessian). This mechanism yields stable RPM trajectories under strong interference and enables recovery from octave ambiguity without ad hoc rules.
\end{itemize}

\section{Problem Formulation}
\label{sec:problem_formulation}

Let $x[n]$ be a measured vibration signal sampled at rate $f_s$. We process $x[n]$ using a sliding window of length $N$ and hop size $H$. The $t$-th frame is the vector
\begin{equation}
\mathbf{x}_t \triangleq [x[n_t],\,x[n_t+1],\,\dots,\,x[n_t+N-1]]^\top,\quad n_t=(t-1)H.
\label{eq:frame_def}
\end{equation}
The rotational speed at frame $t$ is denoted by $r_t$ (RPM) and lies within a physical range $\mathcal{R}=[r_{\min},r_{\max}]$.

In a tacho-less setting, the true speed $r_t$ is unobserved. For each frame, we obtain a set of estimator outputs in the form of one-dimensional evidence curves (defined on estimator-native axes such as lag, quefrency, or frequency). Let $\mathcal{I}$ index the set of estimators and let $\mathcal{E}_t^{(i)}$ denote the evidence curve produced by estimator $i$ at frame $t$. We denote the collection of all curves at frame $t$ as $\mathcal{E}_t \triangleq \{\mathcal{E}_t^{(i)}\}_{i\in\mathcal{I}}$.

The goal is to estimate a speed trajectory and its uncertainty without tachometer input. Specifically, we aim to recursively infer the posterior distribution of the speed given the history of observed evidence curves, denoted as $p(r_t \mid \mathcal{E}_{1:t})$. Based on this posterior (computed on an RPM grid as detailed in Sec.~\ref{sec:method}), we output a trajectory estimate $\hat r_t$ together with an uncertainty summary $\hat\sigma_t$ for each frame $t$.

\section{Proposed Method: Alignment-based RPM Estimation with Curvature-adaptive Tracking}
\label{sec:method}

\subsection{Overview}
ARC consists of three core stages:
(i) \textbf{Likelihood construction (C2G):} Convert each estimator curve into an RPM-grid log-likelihood on a shared RPM grid;
(ii) \textbf{Fusion:} Integrate per-estimator evidence in the log domain while preserving multi-modality; and
(iii) \textbf{Bayesian-inspired recursive filtering:} Propagate uncertainty over time using a curvature-informed, state-dependent motion prior to generate stable trajectories. Note that while we adopt Bayesian terminology for clarity, the likelihoods are heuristic constructions; the approach is Bayesian-inspired rather than fully Bayesian.

\subsection{RPM Grid and Notation}
\label{sec:rpm_grid_notation}
We discretize the feasible RPM range $\mathcal{R}=[r_{\min},r_{\max}]$ into a uniform grid of $G$ points:
\begin{equation}
\mathcal{R}_G \triangleq \{r^{(g)}\}_{g=1}^{G},\quad r^{(g)}=r_{\min}+(g-1)\Delta r,
\label{eq:rpm_grid}
\end{equation}
where $\Delta r = (r_{\max} - r_{\min})/(G-1)$ denotes the grid resolution.
All probabilistic quantities used by the tracker are represented on this discrete grid.
We denote the posterior probability mass at frame $t$ as
\begin{equation}
\pi_t(g) \triangleq p(r_t=r^{(g)} \mid \mathcal{E}_{1:t}),
\label{eq:posterior_mass}
\end{equation}
and the predicted (prior) mass before incorporating frame-$t$ observations as $\pi_t^{-}(g)$.

\subsection{Heterogeneous Evidence Curves}
\label{sec:hetero_curves}
We use three representative tacho-less estimators as observation sources: YIN, cepstrum, and comb matching.
Unlike point-estimate pipelines, ARC retains each estimator's full 1-D search curve per frame.

For each estimator $i\in\mathcal{I}$, the observation tuple is
\begin{equation}
\mathcal{E}_t^{(i)} \triangleq \big(\bm{z}_t^{(i)},\ \bm{c}_t^{(i)},\ \psi^{(i)},\ \kappa^{(i)}\big),
\label{eq:curve_obs}
\end{equation}
where $\bm{z}_t^{(i)}$ is the native axis, $\bm{c}_t^{(i)}$ is the raw curve, 
$\psi^{(i)}\in\{\text{Lag},\ \text{Quefrency},\ \text{Hz},\ \text{RPM}\}$ is the axis type,
and $\kappa^{(i)}\in\{+1,-1\}$ denotes score ($+1$) or cost ($-1$).

\subsubsection{YIN (Lag-domain cost curve)}
YIN provides a lag-domain periodicity error curve $\bm{c}_t^{\text{YIN}}=\{d'(\tau)\}$ on $\bm{z}_t^{\text{YIN}}=\{\tau\}$ with $\psi^{\text{YIN}}=\text{Lag}$ and $\kappa^{\text{YIN}}=-1$.
Lag candidates map to RPM via $r=60f_s/\tau$.

\subsubsection{Cepstrum (Quefrency-domain score curve)}
Cepstrum \cite{Peeters2019MSSP_ReviewComparison} provides a quefrency-domain periodicity strength curve $\bm{c}_t^{\text{Cep}}=\{c(\tau)\}$ on $\bm{z}_t^{\text{Cep}}=\{\tau\}$ with $\psi^{\text{Cep}}=\text{Quefrency}$ and $\kappa^{\text{Cep}}=+1$.
Quefrency candidates map to RPM via $r=60f_s/\tau$.

\subsubsection{Comb matching (Frequency-domain score curve)}
Comb matching \cite{Peeters2019MSSP_ReviewComparison} provides a frequency-domain harmonic-alignment score curve $\bm{c}_t^{\text{Comb}}=\{h(f)\}$ on $\bm{z}_t^{\text{Comb}}=\{f\}$ with $\psi^{\text{Comb}}=\text{Hz}$ and $\kappa^{\text{Comb}}=+1$.
Frequency candidates map linearly to RPM via $r=60f$.

\subsection{Likelihood Construction from Heterogeneous Curves}
\label{sec:curve_to_pdf}
To fuse the heterogeneous observations, we convert each tuple $\mathcal{E}_t^{(i)}$ into a grid-based log-likelihood $\log p_i(r^{(g)} \mid \mathcal{E}_t^{(i)})$ on the shared RPM grid $\mathcal{R}_G$.
\emph{Here, $p_i(\cdot)$ denotes a normalized pseudo-likelihood constructed from estimator evidence, rather than a likelihood derived from an explicit generative signal model.}
This process involves domain mapping, robust standardization, and energy-based probabilistic shaping.

\medskip
\noindent\textbf{Step 1: Domain Mapping.}
We first define a mapping function $r = g_i(z)$ that transforms the native axis $z$ (Lag, Quefrency, or Hz) to RPM based on the axis type $\psi^{(i)}$:
\begin{equation}
g_i(z) =
\begin{cases}
60 f_s / z, & \text{if } \psi^{(i)} \in \{\text{Lag}, \text{Quefrency}\},\\
60 \cdot z, & \text{if } \psi^{(i)} = \text{Hz},\\
z, & \text{if } \psi^{(i)} = \text{RPM}.
\end{cases}
\label{eq:axis_map}
\end{equation}

\medskip
\noindent\textbf{Step 2: Robust Standardization and Energy Formulation.}
Raw curve values $\bm{c}_t^{(i)}$ exhibit heterogeneous dynamic ranges and often contain heavy-tailed outliers caused by resonance or broadband disturbances.
Because standard Z-score normalization is highly sensitive to such multi-modal distributions, we employ robust standardization based on the median and interquartile range (IQR). This effectively suppresses outliers while preserving the relative peak structure required for fusion.
Let $c[m]$ denote the $m$-th element of the raw curve vector $\bm{c}_t^{(i)}$.
We compute the normalized value $\tilde{c}[m]$ as:
\begin{equation}
\tilde{c}[m] = \frac{c[m] - \mathrm{median}(\bm{c}_t^{(i)})}{\mathrm{IQR}(\bm{c}_t^{(i)}) + \epsilon},
\label{eq:robust_norm}
\end{equation}
where $\mathrm{median}(\cdot)$ and $\mathrm{IQR}(\cdot)$ denote the median and interquartile range operators, respectively.
To unify costs (where lower is better) and scores (where higher is better), we define the energy function $E[m]$ using the polarity $\kappa^{(i)}$:
\begin{equation}
E[m] = -\kappa^{(i)} \cdot \tilde{c}[m].
\label{eq:energy_def}
\end{equation}
Under this formulation, lower energy always indicates higher compatibility, regardless of the original estimator type.

\medskip
\noindent\textbf{Step 3: Energy-Based Likelihood via Kernel Aggregation.}
 We define a Gibbs-form pseudo-likelihood from the energy, $p \propto \exp(-\beta E)$. Since energies are robustly standardized (median/IQR) in Eq.~\eqref{eq:robust_norm}, we set $\beta=1$ in all experiments and do not tune it.
This formulation converts estimator-specific mismatch energies into 
observation likelihoods without requiring an explicit generative signal model.
Since the mapping $r=g_i(z)$ is generally non-uniform, we aggregate likelihood contributions onto the uniform RPM grid using a kernel density approach.
The unnormalized likelihood on grid point $r^{(g)}$ is:
\begin{equation}
\tilde{p}_i\!\left(r^{(g)} \mid \mathcal{E}_t^{(i)}\right)
\propto
\sum_{m} \exp\!\big(-\beta\, E[m]\big)\,
K\!\left(\frac{r^{(g)} - g_i(z[m])}{h_{\mathrm{bw}}}\right),
\label{eq:kernel_agg}
\end{equation}
where $z[m]$ is the coordinate of the $m$-th point on the native axis $\bm{z}_t^{(i)}$, $K(\cdot)$ is a kernel function (e.g., Gaussian), and $h_{\mathrm{bw}}$ is the kernel bandwidth.
Finally, we normalize over the grid to obtain the observation likelihood:
\begin{align}
p_i\!\left(r^{(g)} \mid \mathcal{E}_t^{(i)}\right)
&= \frac{\tilde{p}_i\!\left(r^{(g)} \mid \mathcal{E}_t^{(i)}\right)}
{\sum_{j=1}^{G} \tilde{p}_i\!\left(r^{(j)} \mid \mathcal{E}_t^{(i)}\right)}, 
\end{align}
Note that this normalized quantity serves as a heuristic observation likelihood; a fully generative treatment is deferred to future work (see Sec.~\ref{sec:conclusion}).
\subsection{Multi-Estimator Fusion}
Given the per-estimator log-likelihoods, we fuse them via a weighted log-linear pool on the shared RPM grid:
\begin{equation}
\mathcal{L}_t(g) \triangleq \prod_{i\in\mathcal{I}} p_i(r^{(g)}\mid \mathcal{E}_t^{(i)})^{\lambda_i},
\label{eq:fused_lik}
\end{equation}
equivalently,
\begin{equation}
\log \mathcal{L}_t(g) = \sum_{i\in\mathcal{I}} \lambda_i \log p_i(r^{(g)}\mid \mathcal{E}_t^{(i)}),
\label{eq:log_fuse}
\end{equation}
where $\lambda_i$ weights the relative reliability of different estimators (default $\lambda_i=1$).

\subsection{Curvature-Informed State-Dependent Motion Prior}
\label{sec:curvature_transition}
To balance steady-state smoothness and rapid adaptation to sudden RPM changes, we employ a state-dependent Gaussian transition model.
Unlike standard random walks with fixed process noise, our transition variance adapts to the local certainty of the estimate:
\begin{equation}
p(r_t=r^{(g)} \mid r_{t-1}=r^{(j)}) = \mathcal{N}\big(r^{(g)};\, r^{(j)},\, \sigma_{m,t}^2(j)\big),
\label{eq:sd_rw_prior}
\end{equation}
where the variance $\sigma_{m,t}^2(j)$ is inversely modulated by the local curvature of the previous \emph{log-posterior} at state $r^{(j)}$.

Let $\ell_{t-1}(j) = \log(\pi_{t-1}(j) + \epsilon)$ denote the log-posterior mass. 
We approximate the local curvature (Hessian) \cite{mackay2003information} using a second-order central finite difference:
\begin{equation}
\ell''_{t-1}(j) \approx \frac{\ell_{t-1}(j+1) - 2\ell_{t-1}(j) + \ell_{t-1}(j-1)}{(\Delta r)^2}.
\label{eq:finite_diff}
\end{equation}
For numerical stability, we pre-smooth $\ell_{t-1}(j)$ with a fixed 3-point moving average before applying the above finite-difference operator.
For boundary indices, we replicate the nearest interior curvature:
\begin{equation}
\ell''_{t-1}(1) \triangleq \ell''_{t-1}(2), \qquad
\ell''_{t-1}(G) \triangleq \ell''_{t-1}(G-1).
\end{equation}
We define a proxy for local precision (sharpness) as $q_{t-1}(j) = \max(0, -\ell''_{t-1}(j))$.
The transition variance is then adapted via a bounded inverse mapping:
\begin{equation}
\sigma_{m,t}^2(j) = \clip\left(\frac{1}{q_{t-1}(j) + \epsilon_c},\ \sigma_{\min}^2,\ \sigma_{\max}^2\right),
\label{eq:sigma_adapt}
\end{equation}
where $\epsilon_c$ prevents division by zero, and $[\sigma_{\min}^2, \sigma_{\max}^2]$ are the physical bounds for frame-to-frame RPM variation.
Mathematically, this mechanism acts as a curvature-based regularizer: it tightens the transition kernel (low $\sigma^2$) when the tracker is locked onto a sharp peak, and widens it (high $\sigma^2$) to re-acquire targets during transient or uncertain phases.

\subsection{Grid-Based Recursive Filtering}
The core tracking loop recursively maintains the posterior probability mass $\pi_t(g) \triangleq p(r_t=r^{(g)} \mid \mathcal{E}_{1:t})$ on the discrete RPM grid. We perform a recursive update inspired by Bayesian filtering, combining the heuristic likelihood with the motion prior. This recursion proceeds in three steps: prediction (diffusion), update (fusion), and state estimation.

\noindent\textbf{(1) Predict (Time Propagation).}
First, we propagate the previous posterior $\pi_{t-1}$ forward using the state-dependent transition model derived in Sec.~\ref{sec:curvature_transition}. This operation diffuses the probability mass based on the local uncertainty $\sigma_{m,t}^2(j)$, yielding the predicted (prior) mass $\pi^-_t$:
\begin{equation}
\pi^-_t(g) = \sum_{j=1}^{G} \pi_{t-1}(j) \, \mathcal{N}\big(r^{(g)};\, r^{(j)},\, \sigma_{m,t}^2(j)\big).
\label{eq:predict_sd}
\end{equation}

\noindent\textbf{(2) Update (Measurement Fusion).}
Next, we incorporate the fused observation likelihood $\mathcal{L}_t(g)$ from Eqs.~(\ref{eq:fused_lik})--(\ref{eq:log_fuse}). To ensure numerical stability and prevent underflow, this multiplication is performed as summation in the log domain:
\begin{align}
\log \tilde{\pi}_t(g) &= \log(\pi^-_t(g) + \epsilon) + \log \mathcal{L}_t(g), \label{eq:update_log} \\
\pi_t(g) &= \frac{\exp(\log \tilde{\pi}_t(g))}{\sum_{k=1}^{G} \exp(\log \tilde{\pi}_t(k))}, \label{eq:update_norm}
\end{align}
where the normalization ensures that the updated posterior mass $\pi_t(g)$ sums to unity.

\noindent\textbf{(3) Estimation.}
Finally, we compute the Maximum A Posteriori (MAP) and Minimum Mean Square Error (MMSE) estimates from the posterior:
\begin{equation}
\hat{r}_t^{\mathrm{MAP}} = \argmax_{r^{(g)}} \pi_t(g), \quad
\hat{r}_t^{\mathrm{MMSE}} = \sum_{g=1}^{G} r^{(g)}\pi_t(g).
\label{eq:estimates}
\end{equation}
The MMSE serves as the primary smoothed trajectory, while the MAP is reserved for analyzing dominant modes or conflicts.
The estimation uncertainty is quantified by the posterior standard deviation:
\begin{equation}
\hat{\sigma}_t = \sqrt{\sum_{g=1}^{G} \pi_t(g)\big(r^{(g)} - \hat{r}_t^{\mathrm{MMSE}}\big)^2}.
\label{eq:uncertainty}
\end{equation}

\subsection{Algorithm Summary}
Algorithm~\ref{alg:bmhbt_adapt} summarizes the full pipeline.
\begin{algorithm}[t]
\caption{ARC: Alignment-based RPM Estimation with Curvature-adaptive Tracking}
\label{alg:bmhbt_adapt}
\footnotesize
\begin{algorithmic}[1]
\Require Frames $\{\mathbf{x}_t\}_{t=1}^{T}$, Estimators $\mathcal{I}$, RPM Grid $\mathcal{R}_G=\{r^{(g)}\}_{g=1}^{G}$
\Ensure Posteriors $\{\pi_t\}_{t=1}^{T}$ and Estimates $\{\hat r_t^{\mathrm{MAP}},\hat r_t^{\mathrm{MMSE}},\hat\sigma_t\}_{t=1}^{T}$

\State Initialize posterior $\pi_0(g) \leftarrow 1/G$ for all $g=1,\dots,G$

\For{$t=1$ to $T$}

    \State \textbf{Stage I \& II: C2G and Multi-Estimator Fusion}
    \For{each estimator $i \in \mathcal{I}$}
        \State Construct observation tuple $\mathcal{E}_t^{(i)}$ \Comment{Eq.~\eqref{eq:curve_obs}}
        \State Compute grid likelihood $p_i(r^{(g)} \mid \mathcal{E}_t^{(i)})$ \Comment{Eqs.~\eqref{eq:axis_map}--\eqref{eq:kernel_agg}}
    \EndFor
    \State Integrate evidence to obtain fused log-likelihood $\log \mathcal{L}_t(g)$ \Comment{Eq.~\eqref{eq:log_fuse}}

    \State \textbf{Stage III: Curvature-Adaptive Recursive Tracking}
    \State Calculate curvature proxy $q_{t-1}(j)$ via finite difference \Comment{Eq.~\eqref{eq:finite_diff}}
    \State Adapt transition variance $\sigma_{m,t}^2(j)$ based on curvature \Comment{Eq.~\eqref{eq:sigma_adapt}}
    
    \State \textbf{Predict:} Propagate posterior to obtain prior $\pi_t^{-}(g)$ \Comment{Eq.~\eqref{eq:predict_sd}}
    \State \textbf{Update:} Incorporate $\mathcal{L}_t(g)$ and normalize to get $\pi_t(g)$ \Comment{Eqs.~\eqref{eq:update_log}--\eqref{eq:update_norm}}

    \State \textbf{Estimation:} Extract $\hat r_t^{\mathrm{MAP}}, \hat r_t^{\mathrm{MMSE}}$ and compute uncertainty $\hat\sigma_t$ \Comment{Eqs.~\eqref{eq:estimates}--\eqref{eq:uncertainty}}

\EndFor
\end{algorithmic}
\end{algorithm}

\section{Experiments}

\subsection{Experimental Goals and Evaluation Protocol}
We evaluate ARC using a question-driven protocol designed to isolate the contributions of (i) multi-estimator fusion, (ii) temporal tracking, and (iii) real-world feasibility without tachometer reference. Specifically, we ask:

\noindent\textbf{(Q1) Fusion:} Can C2G fusion on a shared RPM grid resolve conflicts among heterogeneous estimators and reduce catastrophic (tail) errors?

\noindent\textbf{(Q2) Tracking:} Beyond static fusion, does temporal recursion (predict/update) reduce jitter and failure propagation compared with a framewise (no-prediction) variant?

\noindent\textbf{(Q3) Real-Data Validation:}
On a real vibration-table recording with mixed phenomena, when an independent tachometer reference is available only for evaluation (not used by ARC during inference), does ARC produce an accurate and physically plausible trajectory with interpretable uncertainty?

\medskip
Across both synthetic and real tests, we report accuracy and tail robustness (RMSE and P95). For real data, we additionally report stability metrics computed from successive-frame RPM increments (jitter and max-jump) to assess trajectory smoothness regardless of ground truth availability.
We organize the evidence as follows: (i) a concise synthetic benchmark (Table~\ref{tab:main_results}); (ii) posterior visualizations explaining conflict resolution (Fig.~\ref{fig:fusion_viz}); (iii) a tracking ablation contrasting framewise fusion vs.\ full tracking (Fig.~\ref{fig:ablation_tracking}); and (iv) a real case study demonstrating uncertainty quantification (Fig.~\ref{fig:real_case}).

\subsection{Synthetic Stress Tests (Failure-Mode Coverage)}
\label{sec:stress_tests}
To rigorously validate ARC, we generate synthetic signals by modulating a harmonic series according to a prescribed ground-truth speed trajectory $\{r_t\}$, yielding an instantaneous fundamental frequency $f_0(t)=r_t/60$.
Crucially, rather than applying generic noise, each stress test is designed to isolate a single failure mechanism commonly observed in industrial vibration.
We select four representative scenarios (\textbf{S1--S4}) for the main quantitative benchmark (Table~\ref{tab:main_results}), while reserving the transient step-change scenario (\textbf{S5}) specifically to visualize tracking dynamics in the ablation study.
Table~\ref{tab:stress_tests_def} details the physical origin and difficulty of each scenario.

\begin{table}[t]
\centering
\caption{Synthetic stress tests used in this paper. S1--S4 are used for the main quantitative benchmark; S5 is used for tracking ablation (Fig.~\ref{fig:ablation_tracking}).}
\label{tab:stress_tests_def}
\footnotesize
\setlength{\tabcolsep}{2.5pt}
\renewcommand{\arraystretch}{1.0}
\begin{tabularx}{\columnwidth}{@{}L{0.06\columnwidth}L{0.46\columnwidth}Y@{}}
\toprule
ID & Stress mechanism (Definition) & Typical physical origin \\ \midrule
\textbf{S1} & \textbf{Octave ambiguity} \newline
\emph{Subharmonic strength}: Amplitude ratio of $0.5\times$ component to fundamental. &
Looseness, rub (subharmonics), strong $2\times$ excitation, oil-whirl-like components \\

\textbf{S2} & \textbf{Low signal-to-noise ratio (SNR)} \newline
\emph{SNR}: Additive broadband noise level relative to signal power. &
Weak excitation, poor mounting, background vibration, electromagnetic noise \\

\textbf{S3} & \textbf{Periodic interference} \newline
\emph{Interference level}: Amplitude of non-harmonic component $f_{\mathrm{int}}=\alpha f_0, \alpha\!\notin\!\mathbb{Z}$. &
Gear mesh, blade-pass, electromagnetic (EM) periodicity, spurious spectral peaks \\

\textbf{S4} & \textbf{Inharmonicity} \newline
\emph{Spectral deviation}: Detuning factor $\delta_m$ shifting harmonics to $m f_0(1+\delta_m)$. &
Torsional slip, stiff spectral structures, resonance-dominated transfer paths \\ \midrule

\textbf{S5} & \textbf{Step change} \newline
\emph{Jump magnitude}: Instantaneous RPM step $\Delta r$ between segments. &
Variable Frequency Drive (VFD) speed command step, load transient \\ \bottomrule
\end{tabularx}
\end{table}

\subsection{Real Vibration-Table Case Study}
\label{sec:real_dataset}
To address \textbf{Q3}, we evaluate ARC on a vibration-table recording characterized by realistic broadband noise and structural resonance.
The test is conducted at an approximately constant operating speed around 1500~RPM.
An \emph{independent tachometer} measurement is available and is used \emph{only as a reference for evaluation} but it is \emph{not} provided to any estimator or to ARC during inference, preserving the strict \textit{tacho-less} setting.

We report quantitative errors against the tachometer speed for this case study.
In addition, to keep the evaluation protocol applicable even when a tachometer is unavailable, we also report stability metrics computed from successive-frame increments $\Delta \hat r_t$ (jitter and max-jump; Sec.~\ref{sec:metrics}), which quantify short-term smoothness and detect implausible jumps regardless of ground truth availability.

\subsection{Methods Compared (Baselines and Variants)}
\label{sec:methods_compared}
We compare against three single-estimator baselines, one internal ablation variant to isolate the contribution of temporal recursion, and one additional real-data baseline representing a classical time--frequency trajectory-extraction pipeline.

\noindent\textbf{Single-estimator baselines:} YIN, cepstrum, and comb matching. Each baseline produces a framewise RPM point estimate by selecting the most plausible candidate on its native search axis, without multi-estimator fusion or temporal recursion.

\noindent\textbf{C2G-Fusion (Tracking Ablation):} A fusion-only variant that performs curve-to-grid likelihood construction and multi-estimator fusion on the shared RPM grid (Sec.~\ref{sec:method}), followed by MMSE estimation, but omits temporal prediction and recursion (i.e., no use of Eq.~\eqref{eq:predict_sd}). This isolates the benefit of heterogeneous fusion (C2G + log-pooling) while removing uncertainty-aware temporal propagation.

\noindent\textbf{ARC (Proposed):} The complete pipeline including C2G likelihood construction, heterogeneous fusion, and curvature-adaptive recursive filtering, producing an online RPM trajectory with posterior uncertainty.

\noindent\textbf{Viterbi-STFT (Real-data baseline):} For the real vibration-table case study only, we additionally include a classical time--frequency baseline that extracts a framewise candidate sequence from a short-time Fourier transform (STFT)-based representation and enforces temporal continuity via Viterbi-style dynamic programming.

\subsection{Evaluation Metrics}
\label{sec:metrics}
\noindent\textbf{Accuracy metrics.}
We denote the reference RPM trajectory as $r_t^{\mathrm{ref}}$, which represents the prescribed ground truth in synthetic scenarios and the independent tachometer measurement in real-world cases. We report the following metrics:
\begin{itemize}
    \item \textbf{RMSE (Root Mean Square Error):} 
    \begin{equation}
        \mathrm{RMSE}=\sqrt{\frac{1}{T}\sum_{t=1}^{T}\big(\hat r_t-r_t^{\mathrm{ref}}\big)^2}.
    \end{equation}
    \item \textbf{P95 (95th-Percentile Absolute Error):} 
    \begin{equation}
    \mathrm{P95}=\text{percentile}_{95}\!\left(\left\{|\hat r_t-r_t^{\mathrm{ref}}|\right\}_{t=1}^{T}\right).
    \end{equation}
\end{itemize}

\textbf{Stability metrics.}
To quantify trajectory smoothness without requiring any reference, we compute successive-frame increments
$\Delta \hat r_t=\hat r_t-\hat r_{t-1}$ for $t=2,\dots,T$:
\begin{itemize}
    \item \textbf{Jitter:} $\mathrm{std}(\Delta \hat r_t)$.
    \item \textbf{Max-Jump:} $\max_{t}|\Delta \hat r_t|$, highlighting abrupt, physically implausible jumps.
\end{itemize}

\subsection{Implementation Details}
\label{sec:impl_details}
Unless otherwise stated, synthetic stress tests use $f_s=12{,}800$~Hz, while the real vibration-table recording is sampled at $f_s=8{,}000$~Hz.
All experiments use the same framing parameters (frame length $N=8192$ samples and hop size $H=128$ samples) and apply the curve-to-grid mapping with the corresponding $f_s$ in \eqref{eq:axis_map}.
 Each trial uses a 5-s signal segment, resulting in $T$ frames determined by $N$, $H$, and the signal length.
We use a uniform RPM grid over $[r_{\min},r_{\max}]=[300,4000]$~RPM with resolution $\Delta r=1$~RPM.
Robust standardization in \eqref{eq:robust_norm} uses $\epsilon=10^{-10}$.
For kernel-based likelihood aggregation in \eqref{eq:kernel_agg}, we use a Gaussian kernel with bandwidth $h_{\mathrm{bw}}=0.5$~RPM and a fixed $\beta=1$.
For curvature-adaptive motion modeling in \eqref{eq:sigma_adapt}, we set $\sigma_{\min}=40$~RPM, $\sigma_{\max}=150$~RPM (i.e., $\sigma_{\min}^2=1600$~RPM$^2$, $\sigma_{\max}^2=22500$~RPM$^2$), $\epsilon_c=10^{-12}$.
For conflict-frame selection in visualizations, we automatically choose the frames based on maximum posterior entropy to avoid cherry-picking.

\section{Results and Discussion}
\subsection{Main Quantitative Analysis}
Table~\ref{tab:main_results} reports RMSE and P95 on four representative synthetic stress tests (S1--S4), aggregated over 20 random seeds.
The results highlight a key trade-off: while single estimators act as ``specialists'' that excel when their specific assumptions hold, they exhibit brittleness in other regimes.
For instance, YIN achieves the best tail performance in inharmonicity scenarios (S4, P95=19.5) but degrades significantly in the presence of octave ambiguity (S1, P95=51.3) and periodic interference (S3, P95=59.0).
Similarly, while Comb matching dominates in the low-SNR regime (S2), it is less robust to interference (S3), where ARC reduces the P95 error by approximately 25\% (25.3 $\to$ 18.9).
In contrast, ARC functions as a robust ``generalist''.
It consistently achieves low P95 errors across diverse scenarios, avoiding the catastrophic failures observed in the baselines (e.g., YIN in S3 or cepstrum in S1/S2) and delivering the most reliable performance across diverse failure modes.

\begin{table}[t]
\centering
\caption{Main results on synthetic stress tests (S1--S4). RMSE and P95 are reported in RPM and aggregated over 20 random seeds.}
\label{tab:main_results}
\small
\setlength{\tabcolsep}{2pt}
\renewcommand{\arraystretch}{1.05}
\begin{tabular}{l|cc|cc|cc|cc}
\toprule
Method & \multicolumn{2}{c|}{S1 } & \multicolumn{2}{c|}{S2 } & \multicolumn{2}{c|}{S3 } & \multicolumn{2}{c}{S4 } \\
 & RMSE & P95 & RMSE & P95 & RMSE & P95 & RMSE & P95 \\
\midrule
YIN      & 40.2 & 51.3 & 26.6 & 52.4 & 53.7 & 59.0 & 16.5 & \textbf{19.5} \\
Cepstrum & 213.0 & 330.8 & 174.2 & 326.1 & 115.2 & 270.5 & 91.7 & 226.9 \\
Comb     & \textbf{12.6} & 31.1 & \textbf{8.5} & \textbf{20.5} & 14.8 & 25.3 & \textbf{15.3} & 28.3 \\
ARC      & 14.3 & \textbf{27.5} & 11.5 & 22.2 & \textbf{10.1} & \textbf{18.9} & 16.1 & 22.2 \\
\bottomrule
\end{tabular}
\end{table}

\begin{figure*}[t]
\centering
\includegraphics[width=0.88\linewidth]{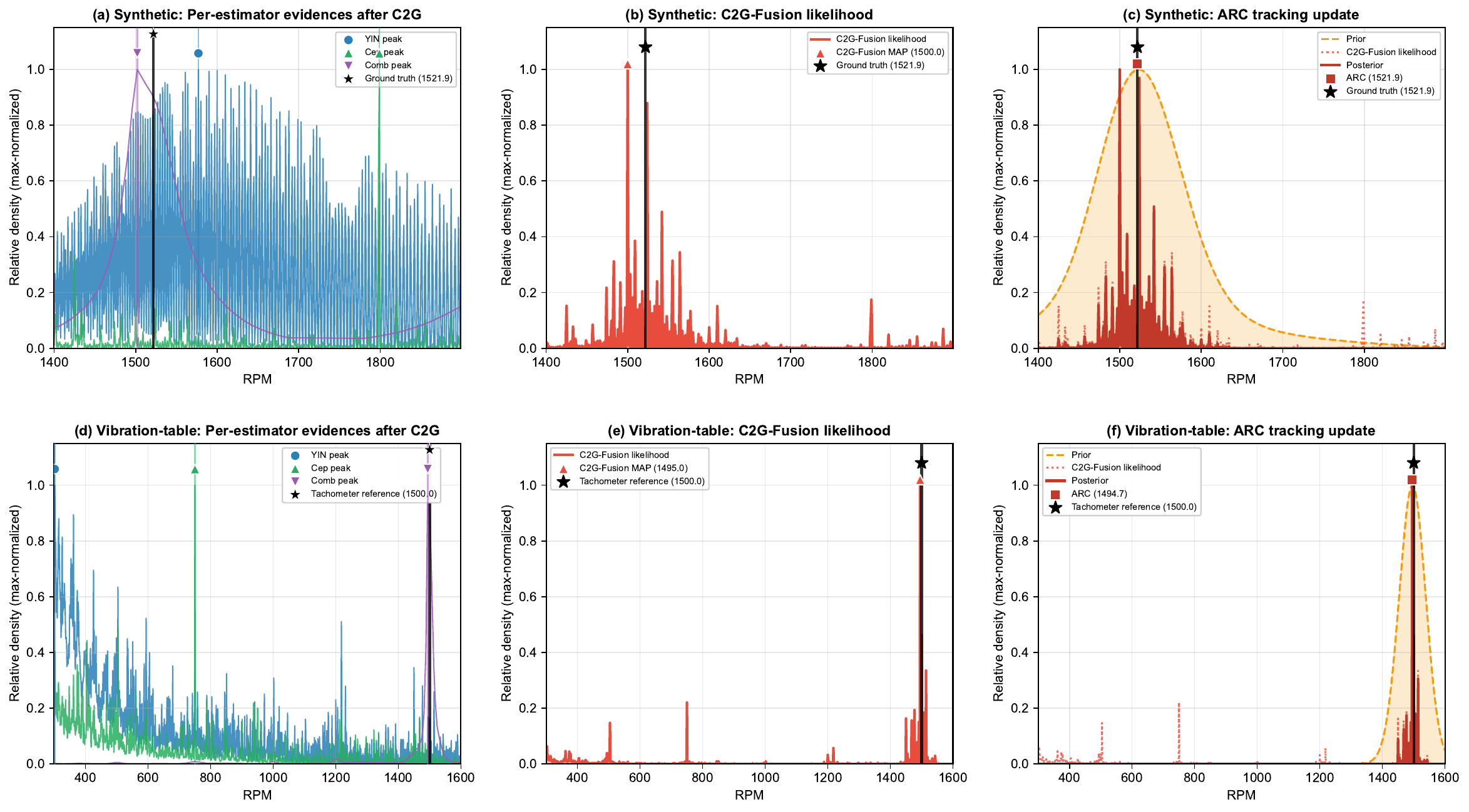}
\caption{%
\textbf{Top row (a--c):} synthetic frame with ground-truth speed.
\textbf{Bottom row (d--f):} vibration-table frame with tachometer reference (evaluation only).
\textbf{(a,d)} Per-estimator C2G evidences on the shared RPM grid.
\textbf{(b,e)} Fused likelihood; C2G-Fusion reports MAP.
\textbf{(c,f)} ARC posterior update; ARC reports MMSE with uncertainty. Curves are max-normalized.
Conflict frames selected by maximum posterior entropy.
}
\label{fig:fusion_viz}
\end{figure*}

\subsection{(Q1) Fusion: Visualization of Conflict Resolution}
\label{sec:results_fusion}

\textbf{Claim:} C2G-based fusion in ARC resolves \emph{structural} disagreement among heterogeneous estimators on a shared RPM grid: it strengthens \emph{mutually supported} hypotheses while suppressing \emph{estimator-specific} outliers, avoiding spurious ``midpoint'' compromises.

Fig.~\ref{fig:fusion_viz} visualizes this mechanism at frames selected by maximum posterior entropy.

\noindent\textbf{Left Column (Incompatibility):}
In (a,d), after C2G, each estimator yields a likelihood-like shape on the \emph{same} RPM axis, yet their dominant modes can be incompatible---for example, periodicity-based cues may favor octave-related candidates, while spectral cues introduce interference-induced peaks. This shows the core difficulty is not a mere scale mismatch but a disagreement over \emph{which RPM regions are plausible}.

\noindent\textbf{Middle Column (Consensus):}
In (b,e), log-domain pooling (Eq.~\eqref{eq:log_fuse}) performs a \emph{consistency-seeking} fusion: since evidence is combined multiplicatively in probability space, only RPM regions that are simultaneously supported by multiple estimators are amplified, whereas isolated peaks are naturally down-weighted. Importantly, this mechanism does not ``average'' peak locations, hence it avoids creating artificial midpoints between incompatible modes.

\noindent\textbf{Right Column (Update):}
In (c,f), the ARC tracking update further integrates the fused evidence with the predicted prior.
Crucially, in the real-data case (f), the resulting posterior effectively resolves the octave conflict seen in (d) and aligns with the tachometer reference.
The posterior shape provides a stable MMSE estimate while retaining residual multi-modality rather than collapsing into an over-confident point decision.

\medskip
Overall, Fig.~\ref{fig:fusion_viz} demonstrates that fusion in ARC acts as a hypothesis filter: it preserves multi-modality when warranted, but removes estimator-specific artifacts without hand-tuned rules.

\subsection{(Q2) Tracking: Benefit Beyond Fusion (Ablation)}
\label{sec:results_tracking}

\textbf{Claim:} Temporal recursion improves robustness beyond per-frame fusion by \emph{rejecting short-lived outliers} and enforcing \emph{physically plausible temporal continuity}, at the cost of a controlled transient lag around abrupt changes.

\begin{figure}[htbp]
\centering
\includegraphics[width=0.92\linewidth]{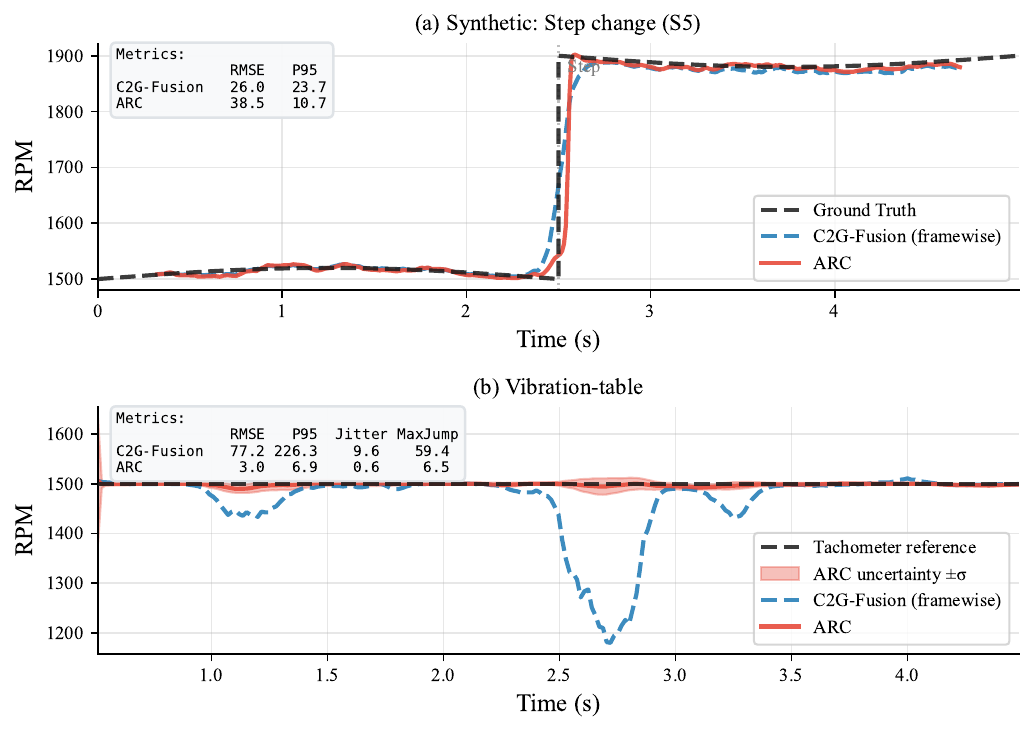}
\caption{%
Tracking ablation: \textbf{C2G-Fusion (Framewise)} vs.\ \textbf{ARC}.
\textbf{(a)} Synthetic step-change (S5) with ground-truth speed.
\textbf{(b)} Vibration-table segment with tachometer reference. Shaded band: $\pm\sigma$.
Inset: RMSE/P95/Jitter/Max-Jump.%
}
\label{fig:ablation_tracking}
\end{figure}

Fig.~\ref{fig:ablation_tracking} compares C2G-Fusion (Framewise) and ARC.
Both variants use the \emph{same} per-frame fusion likelihood on the RPM grid; the only difference is whether we run the \emph{recursive filtering loop}, i.e., applying the prediction in Eq.~\eqref{eq:predict_sd} before the update.
Thus, any performance gap directly reflects the effect of temporal propagation.

\noindent\textbf{(a) Synthetic step-change (S5).}
Around the jump, C2G-Fusion (Framewise) reacts prematurely and spreads error over a longer transition interval.
In contrast, ARC stays on the pre-jump hypothesis until the evidence becomes decisive.
This yields a typical trade-off: the tracker incurs a higher global RMSE (26.0 $\to$ 38.5~RPM) due to the localized lag, yet it suppresses large deviations and reduces the tail error (P95) by over 50\% (23.7 $\to$ 10.7~RPM).
This behavior is expected because RMSE is dominated by a few large squared errors concentrated near the jump, whereas P95 reflects the typical tail deviation over the whole segment.

\noindent\textbf{(b) Vibration-table segment.}
The key failure mode of C2G-Fusion (Framewise) is \emph{failure propagation}: a short-lived conflict triggers a catastrophic dip (see $t\approx 2.7$~s), causing massive errors.
In contrast, ARC leverages the predicted prior to down-weight this transient outlier and remains locked to the previously consistent mode (which aligns with the tachometer reference).
Quantitatively, tracking reduces instability metrics by an order of magnitude (Jitter: 9.6 $\to$ 0.6~RPM; Max-Jump: 59.4 $\to$ 6.5~RPM), and eliminates the massive error caused by the dropout (RMSE: 77.2 $\to$ 3.0~RPM).

\subsection{(Q3) Real Deployment: Industrial-Grade Trajectory \& Confidence}
\label{sec:results_real}

\textbf{Claim:} Under realistic spectral ambiguity where single-cue estimators become unstable, ARC produces a stable RPM trajectory and an interpretable uncertainty signal that reflects evidence coherence, enabling reliable downstream monitoring.

\begin{figure}[htbp]
\centering
\includegraphics[width=\linewidth]{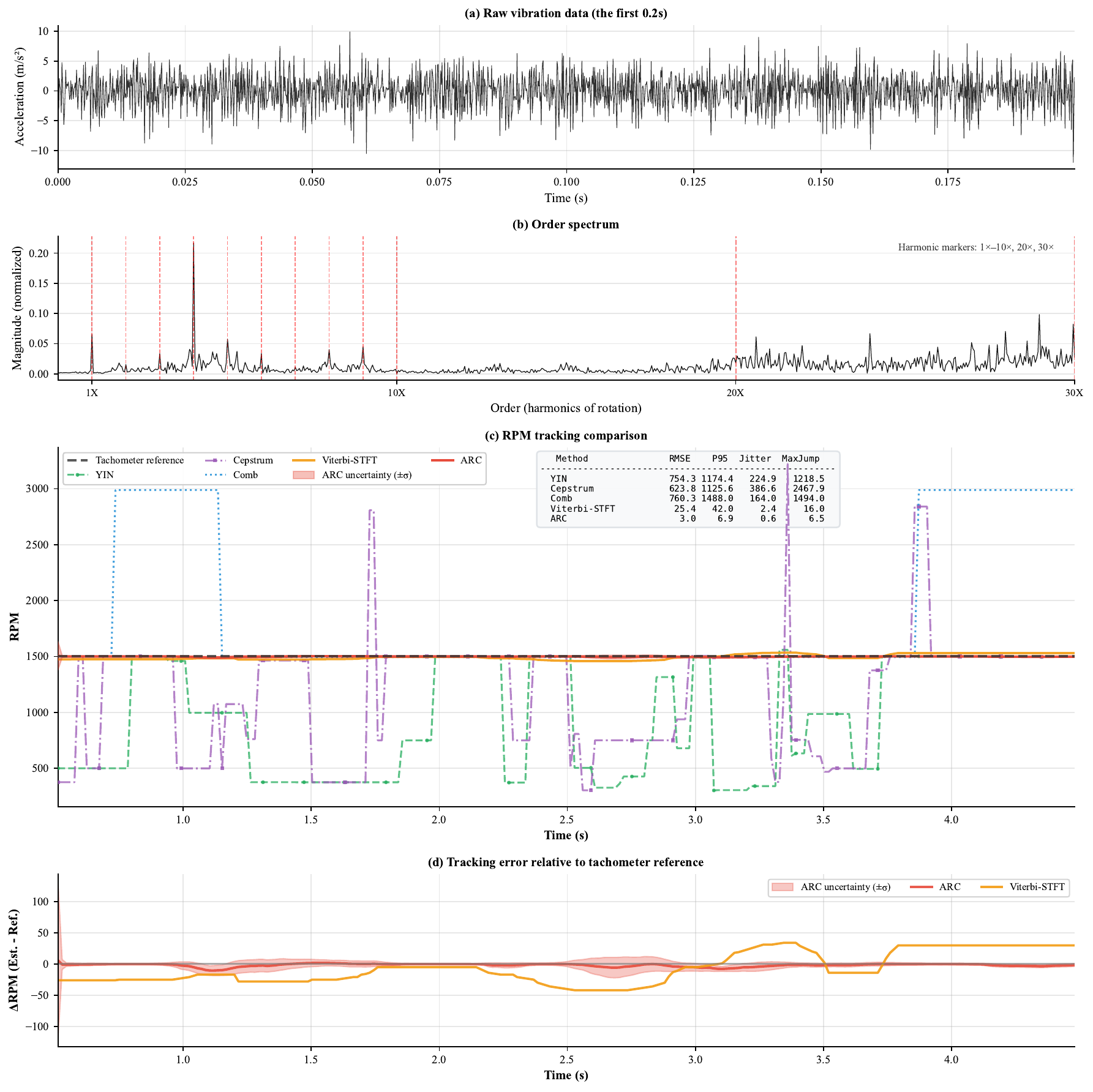}
\caption{%
Real vibration-table case study.
\textbf{(a)} Raw acceleration (0.2\,s).
\textbf{(b)} Order spectrum.
\textbf{(c)} RPM trajectories: YIN/cepstrum/comb (dashed) and Viterbi-STFT vs.\ ARC (MMSE, with $\pm\sigma$ band); inset: metrics.
\textbf{(d)} Tracking error vs.\ tachometer reference.%
}
\label{fig:real_case}
\end{figure}

Fig.~\ref{fig:real_case} evaluates a challenging vibration-table segment. 
A tachometer reference is available for \emph{evaluation only}.

\noindent\textbf{Environmental challenge (a--b):}
Panel~(a) shows a short raw acceleration segment (0.2\,s) with heavy broadband noise and impulsive disturbances.
Panel~(b) shows the order spectrum, where energy is distributed across dense harmonics and interference components rather than being dominated by a single clean 1X peak.
This multi-peak structure creates multiple plausible RPM candidates at the frame level and commonly induces octave-related confusions.

\noindent\textbf{Trajectory robustness (c):}
Panel~(c) compares the \emph{raw framewise} outputs of the constituent estimators (dashed lines: YIN/cepstrum/comb matching) with ARC (red).
The single-estimator traces frequently jump among octave-related or interference-induced candidates, producing physically implausible RPM swings.
In contrast, by pooling evidence on the RPM grid and applying recursive Bayesian-inspired filtering, ARC suppresses short-lived outliers and maintains a coherent trajectory around the 1500\,RPM mode.
Quantitatively, this segment achieves low error and high stability (e.g., RMSE and Jitter reported in the inset table).

\noindent\textbf{Actionable confidence signal:}
The tracker additionally outputs an uncertainty band ($\pm\sigma$, shaded), which acts as a reliability indicator: it broadens during ambiguous/conflicting evidence and contracts when the posterior becomes sharply concentrated.
This enables downstream pipelines to identify low-confidence periods without ad hoc rules.

\noindent\textbf{Error analysis (d):}
Panel~(d) shows the tracking error for both ARC and Viterbi-STFT.
ARC maintains a near-zero error throughout the segment with a tight uncertainty band, while Viterbi-STFT exhibits larger deviations (up to $\pm30$~RPM) during intervals of spectral ambiguity.
This confirms that curvature-adaptive temporal recursion provides more consistent accuracy than trajectory-level dynamic programming.

\section{Conclusion}
\label{sec:conclusion}
We presented a tacho-less speed tracking framework that converts heterogeneous estimator curves into unified grid log-likelihood observations on the RPM axis and performs Bayesian-inspired multi-modal fusion on a discrete RPM grid.
By calibrating evidence directly on the RPM axis (instead of arbitrary metric normalization) and using a curvature-informed state-dependent transition derived from the previous log-posterior (discrete Hessian intuition), the method avoids invalid midpoint averaging, reduces octave-error-induced catastrophes, and removes the need for case-specific jump patching.
The framework is estimator-agnostic: any tacho-less estimator that can provide a per-frame search curve can be integrated by mapping and converting it into an RPM-grid log-likelihood. 

While the posterior uncertainty reported in this work provides a practical confidence indicator, a more rigorous treatment via Bayesian model averaging—explicitly accounting for the possibility that all estimators may be uninformative—could further improve its calibration; this extension will be explored in future work.

\bibliographystyle{IEEEtran}
\bibliography{refs}

\begin{thebibliography}{1}
\providecommand{\url}[1]{#1}
\csname url@samestyle\endcsname
\providecommand{\newblock}{\relax}
\providecommand{\bibinfo}[2]{#2}
\providecommand{\BIBentrySTDinterwordspacing}{\spaceskip=0pt\relax}
\providecommand{\BIBentryALTinterwordstretchfactor}{4}
\providecommand{\BIBentryALTinterwordspacing}{\spaceskip=\fontdimen2\font plus
\BIBentryALTinterwordstretchfactor\fontdimen3\font minus
  \fontdimen4\font\relax}
\providecommand{\BIBforeignlanguage}[2]{{%
\expandafter\ifx\csname l@#1\endcsname\relax
\typeout{** WARNING: IEEEtran.bst: No hyphenation pattern has been}%
\typeout{** loaded for the language `#1'. Using the pattern for}%
\typeout{** the default language instead.}%
\else
\language=\csname l@#1\endcsname
\fi
#2}}
\providecommand{\BIBdecl}{\relax}
\BIBdecl

\bibitem{Lu2019TIM_TacholessReview}
S.~Lu, R.~Yan, Y.~Liu, and Q.~Wang, ``Tacholess speed estimation in order
  tracking: A review with application to rotating machine fault diagnosis,''
  \emph{IEEE Transactions on Instrumentation and Measurement}, vol.~68, no.~7,
  pp. 2315--2332, 2019.

\bibitem{Peeters2019MSSP_ReviewComparison}
C.~Peeters, Q.~Lecl{\`e}re, J.~Antoni, P.~Lindahl, J.~Donnal, S.~Leeb, and
  J.~Helsen, ``Review and comparison of tacholess instantaneous speed
  estimation methods on experimental vibration data,'' \emph{Mechanical Systems
  and Signal Processing}, vol. 129, pp. 407--436, 2019.

\bibitem{deCheveigne2002YIN}
A.~de~Cheveign{\'e} and H.~Kawahara, ``Yin, a fundamental frequency estimator
  for speech and music,'' \emph{The Journal of the Acoustical Society of
  America}, vol. 111, no.~4, pp. 1917--1930, 2002.

\bibitem{Peng2023MSSP_AssessImprove}
D.~Peng, Y.~Chen, M.~J. Zuo, and C.~K. Mechefske, ``Assessment and improvement
  of the accuracy of tacholess instantaneous speed estimation,''
  \emph{Mechanical Systems and Signal Processing}, vol. 202, p. 110706, 2023.

\bibitem{Yoo2024MSSP_WMOVA}
J.~Yoo, J.~Park, T.~Kim, J.~M. Ha, and B.~D. Youn, ``Weighted multi-order
  viterbi algorithm (wmova): Instantaneous angular speed estimation under harsh
  conditions,'' \emph{Mechanical Systems and Signal Processing}, vol. 211, p.
  111187, 2024.

\bibitem{Feng2023SHM_VKF}
K.~Feng, J.~C. Ji, and Q.~Ni, ``A novel adaptive bandwidth selection method for
  vold--kalman filtering and its application in wind turbine planetary gearbox
  diagnostics,'' \emph{Structural Health Monitoring}, vol.~22, no.~2, pp.
  1027--1048, 2023.

\bibitem{Zhang2025MST_ARE}
D.~Zhang, Z.~Idrus, and R.~Hamzah, ``An automatic ridge estimation method for
  tacholess order tracking of wind turbine gearbox under varying-speed
  operation,'' \emph{Measurement Science and Technology}, vol.~36, no.~8, 2025.

\bibitem{Leclere2016MOPA}
Q.~Lecl{\`e}re, H.~Andr{\'e}, and J.~Antoni, ``A multi-order probabilistic
  approach for instantaneous angular speed tracking: Debriefing of the cmmno'14
  diagnosis contest,'' \emph{Mechanical Systems and Signal Processing},
  vol.~81, pp. 375--386, 2016.

\bibitem{mackay2003information}
D.~J.~C. MacKay, \emph{Information Theory, Inference and Learning
  Algorithms}.\hskip 1em plus 0.5em minus 0.4em\relax Cambridge: Cambridge
  University Press, 2003.

\end{thebibliography}
\end{document}